# Ultra-Low Insertion Loss HTSC S-Band Stepped-Impedance Resonator Filter for RF Front-Ends

**Ilan Kurtser[1], Yoav Koral[1], Eldad Holdengreber[2,*] Shmuel E. Schacham[1], Eliyahu Farber[1,3]**

[1] Department of Electrical & Electronic Engineering, Ariel University, ISRAEL
[2] Department of Mechanical Engineering & Mechatronics, Ariel University, ISRAEL
[3] Department of Physics, Ariel University, ISRAEL

*Author to whom any correspondence should be addressed.

E-mail: eldadh@ariel.ac.il

## Abstract

We present the design, fabrication, and measurement of a high-temperature superconductor (HTSC) Stepped Impedance Resonator (SIR) band-pass filter for S-band applications, and its incorporation into a cryogenic receiver cascade. The 11-pole filter, implemented in $YBa_2Cu_3O_{7-x}$ (YBCO) thin films on sapphire, exhibits an ultra-low insertion loss (IL) of −0.1 dB, a sharp roll-off of 100 MHz, and a rejection level exceeding −80 dB. These measured results represent, to the best of our knowledge, the lowest reported IL for an S-band filter with this number of poles. When integrated with a cryogenic low-noise amplifier (LNA), system-level simulations and measurements predict a receiver noise figure (NF) of 0.34 dB at 3.39 GHz, enabling a ~20% increase in radar detection range compared with conventional copper-based front ends. This work demonstrates the feasibility of practical HTSC-based RF front-ends for next-generation communication and radar systems.

## 1. Introduction

Improving radar detection range can be achieved either by increasing transmitted power or by lowering the noise figure (NF) of the receiver. Conventional receiver frontends, based on copper microstrip filters and semiconductor LNAs, are typically limited by insertion losses of several dB in the passband, which directly increases the overall NF.

High-temperature superconductors (HTSCs), such as YBCO, offer exceptionally low surface resistance at microwave frequencies when cooled with liquid nitrogen, making them ideal candidates for RF front-end applications [1–11]. Previous studies have shown HTSC-based devices at L-band and VHF with improved performance [12–16]. However, experimental demonstrations of multi-pole S-band HTSC band-pass filters integrated into front-end cascades remain limited.

In this work, we extend our earlier simulation study [9] by reporting the first measured demonstration of an 11-pole SIR HTSC filter in the S-band, followed by integration into a cryogenic front-end cascade with an LNA.

We show both component-level and system-level improvements in insertion loss, noise performance, and radar detection range in comparsion to conventional components.

This work targets the S-band, a key frequency range for communication and radar, through the design of an 11-pole Stepped Impedance Resonator (SIR) band-pass filter with an in-band insertion loss of about −0.1 dB. At the system level, simulations of a receiver front-end operating at 77 K—combining two such filters with a cryogenic low-noise amplifier—predict an ultra-low noise figure, highlighting the potential of HTSC technology to significantly improve receiver sensitivity.

## 2. Methods

### 2.1 Filter Design and Simulation

The filter was designed in Cadence AWR, using an 11-pole SIR microstrip topology. Compared with hairpin or edge-coupled topologies, the SIR structure offers a wider rejection bandwidth and lower in-band insertion loss at the cost of larger physical dimensions [17].

The SIR structure comprises resonators with varying impedances, consisting of two wide capacitive transmission lines and a thin inductive transmission line, adjusted according to the required resonance frequency

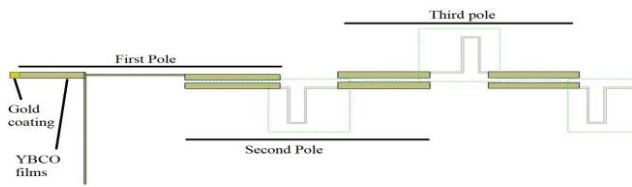

**Figure 1.** First 3 poles SIR microstrip geometry of the YBCO filter

To reduce the resonator's dimensions, the inductive section is bent at four points, thus creating a "u" shaped transmission line while ensuring that the transmission line remains sufficiently narrow to prevent additional coupling at the bends, as shown enclosed in green dotted squares in Figure 1. Additional dimensions of length, coupling lines, distance, and width are presented in the Appendix.

The cross-section of the YBCO microstrip filter is shown in Figure 2. The dimensions are not to scale.

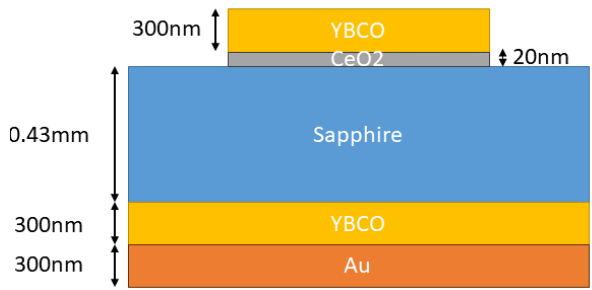

**Figure 2.** Cross-section of the YBCO-based microstrip filter.

To investigate the performance of the superconducting filter, the filter was simulated as a perfect conductor [18-22]. To reduce simulation time, the HTSC-based conductor thin film microstrips were initially simulated as zero-thickness layers [13], thus reducing mesh complexity and allowing faster simulations. After optimizing the resonators' width, length, and coupling distance, the final SIR geometry was simulated with a thick 300 nm conductor. The final dimensions of the filter are 84 mm long and 14.2 mm wide.

The simulation results of IL ($S_{21}$, pink squares) and return loss ($S_{11}$, blue triangles) as a function of frequency are shown in Figure 3. The results show ultra-low losses at the filter passband, with a loss of -0.068 dB at 3.45 GHz. This result demonstrates the potential high performance of devices implemented using HTSC, significantly better than those obtained through simulations of conventional filters, in which typical IL is about 3 dB in the passband [17].

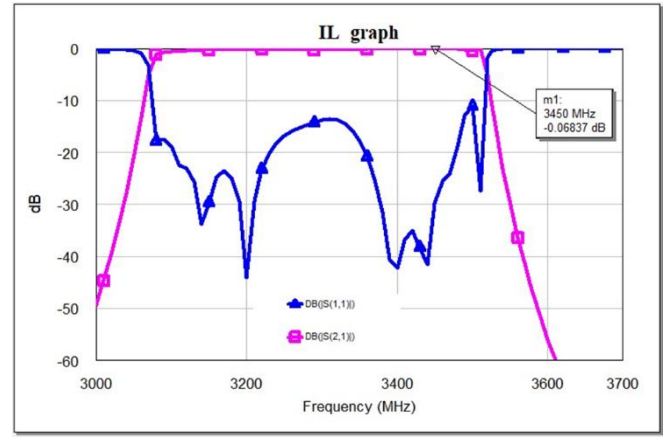

**Figure 3.** Simulation of IL S21 (purple squares) and Return Loss S11 (blue triangles) vs. frequency of HTSC filter.

The SIR is a geometry optimal for RF frontends by displaying a high rejection at the rejection band and is constructed using parallel coupled resonators [14, 23-25]. Each resonator structure is made of a wide capacitive part and a thin inductive part, thus creating resonance for a specific frequency. To shorten the filter's length the inductive part was bent in four points, without compromising its performance while keeping its original length [7].

### 2.2 Filter Fabrication and Measurement

The filters were fabricated. based on simulation results. The 0.43 mm thick sapphire substrate was coated on both sides with 300 nm YBCO layers, separated by a 20 nm $CeO_2$ buffer layer to facilitate epitaxial growth. The top layer was milled with dry lithography, and the bottom layer was coated with another 300 nm layer of Au. Wide gold-coated pads were added at the filter's edges to enable ultrasonic bond connections.

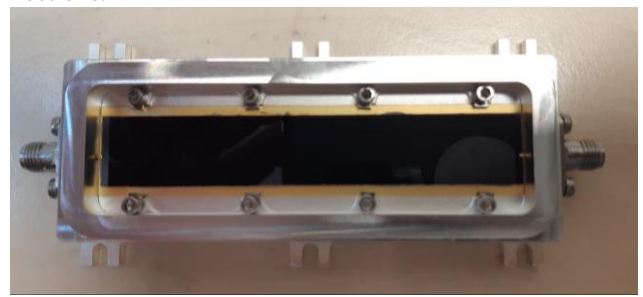

**Figure 4.** Filter in casing before laser sealing.

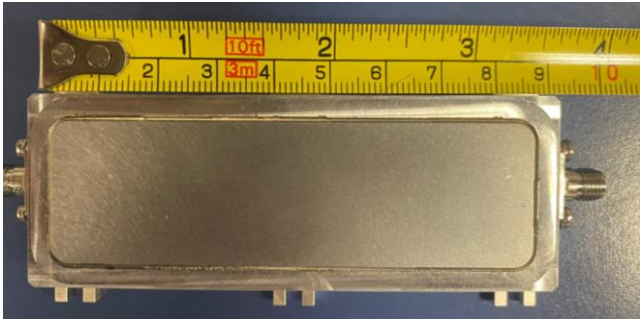

**Figure 5.** Filter in casing after laser sealing.

The superconducting filter's S-Parameters were measured using an Electromagnetic Network Analyzer (ENA). The filter was cooled below the $T_C$ of YBCO using liquid nitrogen. The cooling cannot be achieved by submerging the bare filter in the liquid without risking damage to the device. Corrosion from icing or cracks of the material resulting from thermal shock can seriously deteriorate the filter's performance [26]. A mechanical casing was designed to isolate the filter from the liquid nitrogen to prevent damage. The casing was constructed from aluminum due to its high heat conduction. A CuMo (Copper-Molybdenum) carrier was incorporated to manage aluminum and gold's thermal expansion and contraction (Figure 4). To prevent icing on the filter, the casing was laser-sealed with inert gas enclosed within the chamber, eliminating atmospheric humidity (Figure 5).

**שגיאה! מקור ההפניה לא נמצא.**6 shows the filter return loss $S_{11}$ (blue line) and IL $S_{21}$ (red line). The measured results agree closely with the simulation results depicted in **שגיאה! מקור ההפניה לא נמצא.**3. The filter exhibits a steep passband slope, with a roll-off of 100 MHz on both sides, transitioning from a rejection band of approximately -82 dB to a passband around -0.1 dB, with certain frequencies showing even lower IL. While the IL data is impressive, a notable shortcoming is the high return loss $S_{11}$, approximately -20 dB. This compromise was made in the design stage to achieve better insertion loss. Nonetheless, these results demonstrate the high performance of the superconducting filter. The data confirms excellent selectivity, low IL, and consistent operation at cryogenic temperatures, indicating its suitability for high-sensitivity applications.

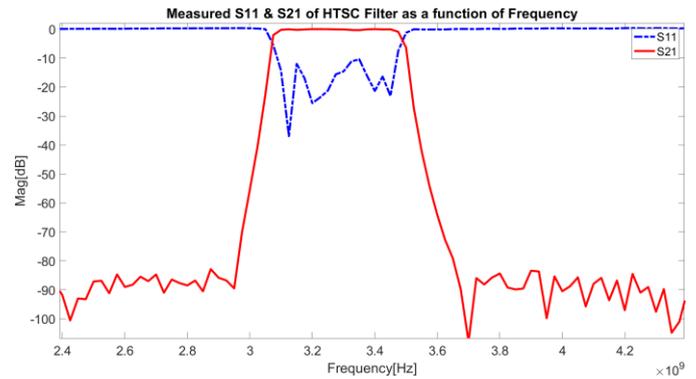

**Figure 6.** Measured results of IL $S_{21}$ (red line) and Return Loss $S_{11}$ (blue dashed line) of HTSC filter vs. frequency.

The $S_{21}$ simulation results for the HTSC filter, obtained using the AWR software, modeled as a zero thickness Perfect Electric Conductor (blue diamond), and the 300nm thick HTSC device $S_{21}$ measured data (orange squares) are presented in Figure 7. The figure shows the excellent fit between the AWR simulation results and the experimental data (orange squares). This level of agreement between AWR simulations and measurements is consistent with previous RF microstrip designs [27].

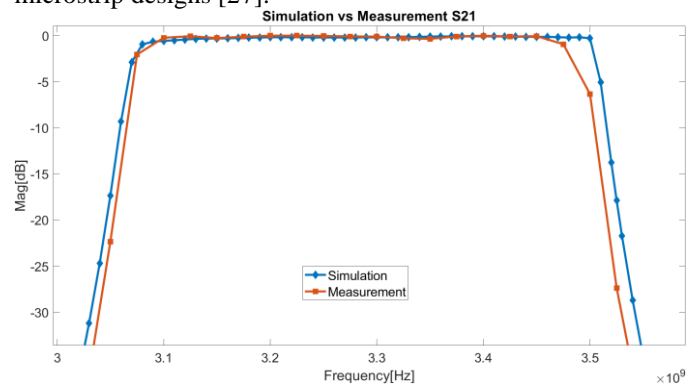

**Figure 7.** $S_{21}$ simulation result (blue diamonds) and measurement (orange squares) vs. frequency.

## 3. Results

### 3.1 Front-End Cascade Simulation and Measurement

To evaluate the noise performance, a noise figure simulation of the complete front-end cascade was carried out, consisting of the HTSC filter followed by a commercial cryogenic low-noise amplifier (LNA) [28-31]. The simulation results were further compared with experimental measurements to assess the impact of the HTSC filter on overall receiver performance.

System-level simulations were carried out using Friis' formula for cascaded noise figure. Based on the simulated filter insertion loss of –0.07 dB and the LNA's datasheet parameters (gain ≈ 30 dB, NF ≈ 0.05 dB at 77 K), the cascade is expected to achieve an effective NF of 0.057 dB at 3.3 GHz.

The predicted NF response across the passband is shown in Figure 8, remaining nearly flat and below 0.1 dB within the 3.1–3.5 GHz band.

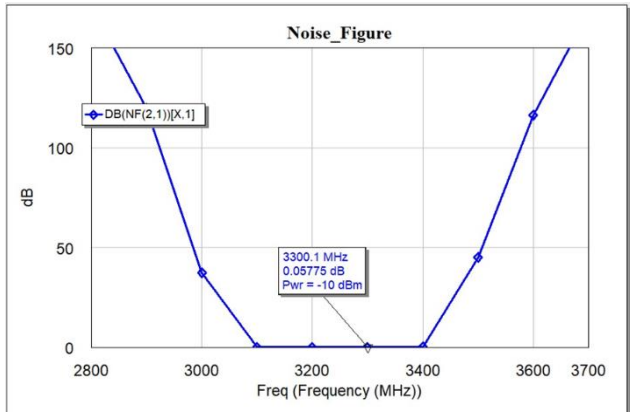

**Figure 8.** Noise Figure simulation of HTSC front-end cascade as function of frequency.

The filter and LNA were submerged in liquid-nitrogen dewar and characterized using a calibrated noise source and a noise figure analyzer. Figure 9 shows the measured NF versus frequency (3.05–3.5 GHz).

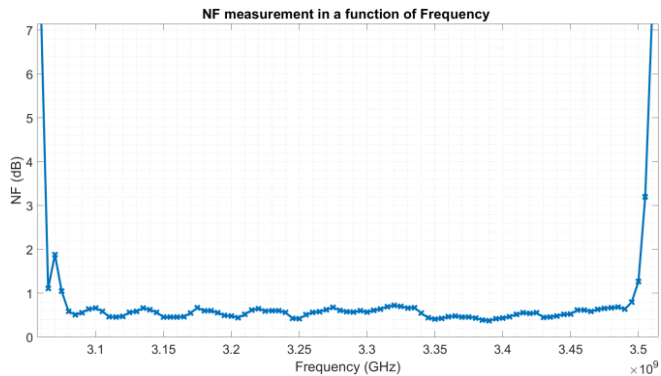

**Figure 9.** Measured NF of HTSC front-end cascade as a function of frequency.

The measured results, shown in Figure 9, indicate that the cascade maintains a noise figure below 0.6 dB across most of the passband, with a minimum value of 0.37 dB occurring at 3.39 GHz. Outside the passband, the NF increases sharply, consistent with the filter's high rejection characteristics. These results confirm that the combination of the HTSC filter and cryogenic LNA achieves exceptionally low-noise performance in the S-band.

The measured NF is higher than the simulated value, primarily due to residual insertion losses from connectors, cryostat feedthroughs, and imperfect impedance matching. Nevertheless, the measured NF is still an order of magnitude lower than conventional copper-based front-ends (~3–4 dB). This demonstrates that the HTSC filter substantially improves overall receiver sensitivity, much superior to that of commercial Frontend [29].

## Discussion

The results clearly demonstrate that HTSC filters provide a path toward ultra-low noise receiver front-ends. The measured filter IL of −0.1 dB is unprecedented in the S-band for an 11-pole design. The sharp roll-off and high rejection make it particularly suitable for congested spectral environments such as 5G, satellite ground stations, and radar systems.

Based on the Power Density Equation, we get the maximum range equation (1)[32]:

$$R_{Max} = \left( \frac{P_t G^2 \lambda^2 \sigma}{(4\pi)^3 k T_0 B F (SNR)_{0_{Min}}} \right)^{\frac{1}{4}} \qquad (1)$$

where $P_t$ is the transmitted power, $G$ is the gain of the directional antenna, $\lambda$ is the operational wavelength, $\sigma$ is the target's radar cross section, $B$ is the receiver bandwidth, $k$ is Boltzmann's constant, $T_0$ is the temperature in Kelvin, $SNR_{min}$ is the minimum Signal-to-Noise ratio for detection, and $F$ is the NF on a linear scale, i.e.

$$F \equiv 10^{NF/10}$$

Inserting the results of Figure 6b, with the help of (2), into the maximum range equation (1) of a radar receiver and comparing the results to the NF for a conventional conductor frontend with a typical NF of 4 dB, yields a range improvement of around 20%.

## Conclusion

We have presented the design, fabrication, and experimental characterization of an 11-pole S-band band-pass filter based on high-temperature superconducting YBCO thin films, and its integration into a cryogenic receiver front-end. The filter demonstrates an exceptionally low insertion loss of approximately −0.1 dB, a steep 100 MHz roll-off, and a rejection level exceeding −80 dB, representing one of the best reported performances for this frequency range and filter order.

When incorporated into a cascade with a cryogenic LNA, the system achieves a measured minimum noise figure of 0.37 dB at 3.39 GHz, in close alignment with simulation trends that predicted a near-ideal NF of 0.057 dB. The discrepancy between measured and simulated values is attributed mainly to connector, packaging, and cryostat losses, yet the measured

NF remains significantly superior to conventional copper-based front ends, which typically exhibit 3–4 dB NF [7,31].

These results demonstrate the feasibility and practical advantages of employing HTSC filters in next-generation radar and communication systems, where ultra-low noise and high selectivity are critical [33]. Future work will focus on improving impedance matching to further reduce return loss, scaling the filter design to broader bandwidths, and validating performance in a full operational radar environment.

## Acknowledgements

The authors would like to thank Mr. Benny Hadad of Cadence Design Systems for fruitful discussions.

## Conflict of Interest

The authors declare no competing financial interests or personal relationships that could have appeared to influence the work reported in this paper.

## Data Availability Statement

The data that support the findings of this study are available from the corresponding author upon reasonable request.

## Ethics Statement

This study did not involve human participants, animals, or personally identifiable data. Ethics approval was not required.